\documentstyle[aas2pp4]{article}

\lefthead{Brandner \& K\"ohler}
\righthead{The distribution of pre-main-sequence binary separations}

\begin{document}

\title{Star formation environments and the distribution of 
binary separations\footnote{Based on observations obtained at the European 
Southern Observatory, La Silla; ESO Proposal No.\ 53.7-0121, 55.E-0968, 56.MPI-8}}

\author{Wolfgang Brandner\altaffilmark{2,3}, Rainer K\"ohler
\altaffilmark{4,5}}
\affil{$^2$Caltech - JPL/IPAC, Mail Code 100-22, Pasadena, CA 91125, USA}
\affil{$^3$University of Illinois at Urbana-Champaign, Department of Astronomy,
1002 West Green Street, Urbana, IL 61801, USA}
\authoremail{brandner@ipac.caltech.edu}

\affil{$^4$Max-Planck-Institut f\"ur Astronomie, K\"onigstuhl 17, D-69117 
Heidelberg, Germany}
\affil{$^5$Astrophysikalisches Institut Potsdam, An der Sternwarte 16,
D-14482 Potsdam, Germany}
\authoremail{koehler@aip.de}

\begin{abstract} 
We have carried out K-band speckle observations of a sample of 114 X-ray 
selected weak-line T\,Tauri stars in the nearby Scorpius-Centaurus OB 
association. We find that for binary T\,Tauri stars closely associated to the 
early type stars in Upper Scorpius, the youngest subgroup of the OB 
association, the peak in the distribution of binary separations is at 90 A.U. 
For binary T\,Tauri stars located in the direction of an older subgroup, but not
closely associated to early type stars, the peak in the distribution is at 215 
A.U. A Kolmogorov-Smirnov test indicates that the two binary populations do not
result from the same distribution at a significance level of 98\%.

Apparently, the same physical conditions which facilitate the formation of 
massive stars also facilitate the formation of closer binaries among low-mass
stars, whereas physical conditions unfavorable for the formation of massive 
stars lead to the formation of wider binaries among low-mass stars. The outcome
of the binary formation process might be related to the internal turbulence and
the angular momentum of molecular cloud cores, magnetic field, the initial 
temperature within a cloud, or - most likely - a combination of all of these. 
 
We conclude that the distribution of binary separations is not a universal 
quantity, and that the broad distribution of binary separations observed among 
main-sequence stars can be explained by a superposition of more peaked binary 
distributions resulting from various star forming environments. The overall 
binary frequency among pre-main-sequence stars in individual star forming 
regions is not necessarily higher than among main-sequence stars.

\end{abstract}


\section{Introduction}

Taurus-Auriga is the star forming region which has been most thoroughly surveyed
for pre-main-sequence binary and multiple systems (see Mathieu 1994 for a 
review). For separations from 15 A.U.\ to 
2000 A.U., the binary frequency among T\,Tauri stars in Taurus is 1.9 times 
as high as among nearby main-sequence stars (K\"ohler \& Leinert 1998). 
Extrapolating over the whole range of separations yields
a binary frequency of 100\%, i.\,e., each T\,Tauri star in Taurus should be 
member of a binary or multiple systems. This apparent overabundance of
binaries among pre-main-sequence stars is puzzling. 
One possible explanation is a decrease in the binary frequency as a function
of the age of a stellar population (Patience et al.\ 1998).
However, a T\,association like Taurus 
might not be the typical birthplace for low-mass stars, as up to 80\% of 
all low-mass stars could originate in OB associations (Miller \& Scalo 1978;
see also Zinnecker et al.\ 1992).

Scorpius-Centaurus is the most nearby OB association at a distance of about 145
parsec (de Zeeuw et al.\ 1998). It consists of three subgroups 
(cf.\ Figure 1) with ages ranging from 5 to 13 Myr (de Geus et al.\ 1989). 
Upper Centaurus-Lupus (UCL) is the oldest subgroup of
the association. Star formation started here 13 Myr ago and
subsequently progressed throughout the parental giant molecular cloud
(e.g.\ Blaauw 1991 and references therein).

Based on observations with the EINSTEIN X-ray satellite, Walter et al.\ (1994)
identified 28 weak-line T\,Tauri stars (WTTS) in Upper Scorpius (US), the 
youngest subgroup. 10 of these
have been surveyed by Ghez et al.\ (1993) for binary or multiple systems,
and three binaries have been detected.
The EINSTEIN fields covered only a small fraction of US (Fig.\ 2)
and the 28 WTTS did not allow for a statistical meaningful study of
binary frequencies and separations. A search for visual binary stars among 
74 ROSAT selected WTTS and post-T\,Tauri stars in US (Kunkel et al., in prep) 
was carried by Brandner et al.\ (1996). This survey was sensitive to binary
separations down to 0\farcs8, and revealed a rather high binary
frequency in the region located between US and UCL (`US-B') and an
apparent absence of wide visual binary stars in the center of US (`US-A').

In order to identify closer binary systems and to get a better statistics
on possible spatial variations of binary star properties, we have carried
out a speckle survey of 114 WTTS in US based on the
lists by Walter et al.\ (1994) and Kunkel et al.\ (in prep).

%
% Fig. 1
%
\begin{figure*}[htb]
\centerline{\plotfiddle{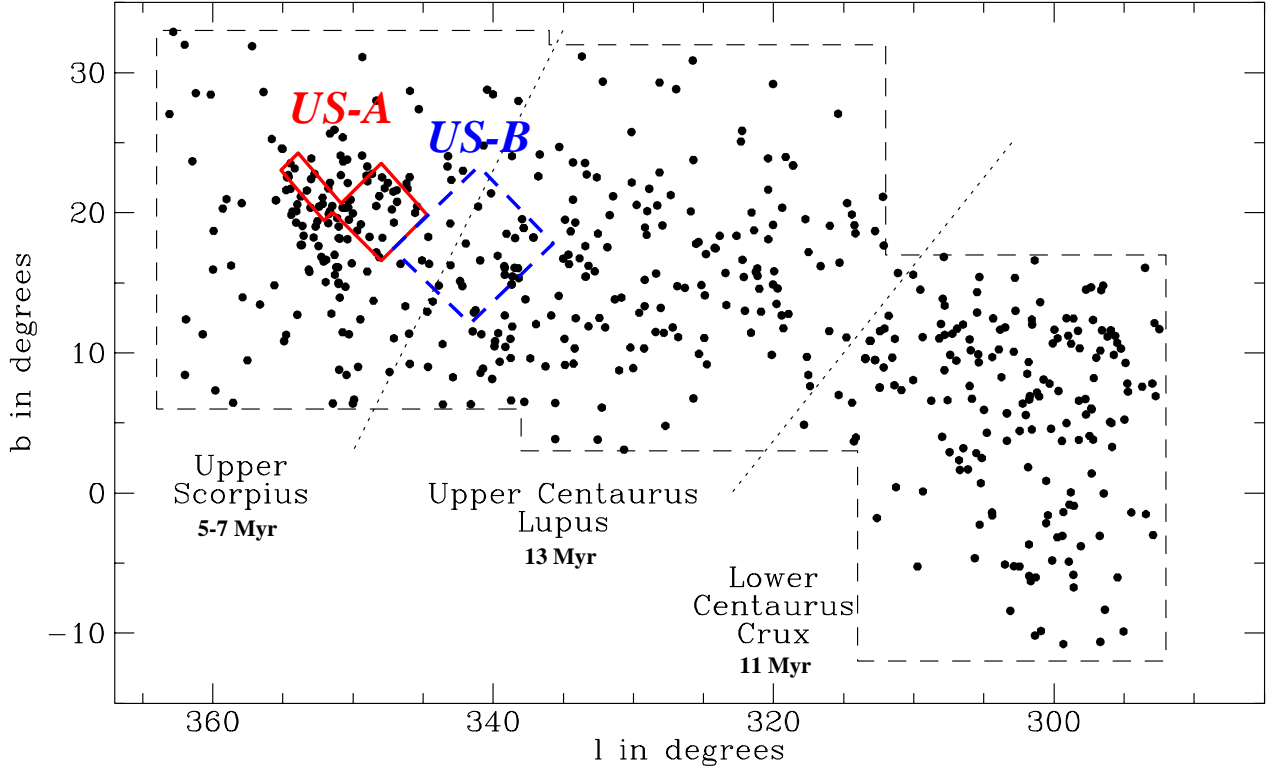}{11cm}{0}{90}{90}{-520}{-310}}
\figcaption{The spatial distribution of 532 proper motion members of the
Scorpius-Centaurus  association based on HIPPARCOS measurements (adapted
from de Zeeuw
et al.\ 1998). The boundaries between the subgroups Upper Scorpius (US), Upper
Centaurus Lupus (UCL), and Lower Centaurus Crux (LCC) are indicated by dotted
lines.  Star formation has progressed from the oldest subgroup UCL
towards the younger subgroups US and LCC. The two adjacent fields of our
survey for binary T\,Tauri stars, centered on US (US-A, solid lines) and at
the interface between US and UCL (US-B, dashed lines) are outlined.
\label{fig1}}
\end{figure*}

\section{Observations and Data Analysis} 

The speckle data were obtained with the SHARP-camera at the ESO 3.5m New 
Technology Telescope on La Silla, Chile, in May 1994 and July 1995.
All observations were performed in the K-band.
In order to estimate the local density of background sources, we obtained
additional infrared images with the ESO/MPIA $2.2\rm\,m$ telescope on
La Silla in March 1996 using the IRAC2b camera.
The detailed results, i.\,e., separations, position angles, photometry of the 
individual binary components, and de-biased binary frequencies will be presented
in a forthcoming paper (K\"ohler et al., in prep). 

Here we concentrate
on the distribution of the binary separations.
In total we observed 68 WTTS in US-A (26 from the list by Walter et al.\ 1994,
and 42 new WTTS from Kunkel et al., in prep) and 46 WTTS in US-B (Kunkel et 
al., in prep). As US is projected onto the Galactic bulge and in the K-band 
has a 6$\times$ higher density of background sources than Taurus 
(K\"ohler 1997; K\"ohler et al., in prep.), we used a separation of 3$''$ as 
the upper 
cut-off, in order to reduce contamination by background giants.
The distribution of brightness ratios as a function of binary
separations is rather uniform for separations between 0\farcs1 and 3$''$. 
The limiting magnitude for undetected companions has a 1/d dependence
with a typical brightness ratio limit of 0.16 at 0\farcs5 and 0.1 at 0\farcs8.
Among the binaries with separations between 0\farcs8 and 3$''$, only two have
brightness ratios $\le$~0.1. For separations $\ge$~4$''$
we find a larger number of visual binaries with brightness ratios $\le$~0.1,
most of which are very likely chance projections. 
Counting binaries with separations between 0\farcs1 and 3$''$,
we detect 21 binaries in US-A and 18 binaries (if we count one hierarchical
triple system as two binaries) in US-B.

\section{Comparison of US-A and US-B}
 
The age of the WTTS population in US is under debate. Walter et al.\ (1994)
derive ages of 1-2 Myr for the WTTS and conclude that the WTTS formed 3-4 Myr
{\it after} the early-type stars in US.
Kunkel et al.\ (in prep) and Mart\'{\i}n (1998) find that the
age of WTTS in Upper Scorpius is in agreement with the age of the B 
stars. Our speckle survey uncovers that almost all of the brighter M-type WTTS
in US-A and US-B are in fact binaries.
Only if they are not recognized as such, they are erroneously placed too high
in the H-R diagram (cf.\ Simon et al.\ 1993, Brandner \& Zinnecker 1997,
Ghez et al.\ 1997) and an age of
$\le$ 1 Myr is derived.
% If low-mass and high-mass stars formed at the same
% time, the physical conditions in the molecular cloud at that time also
% determined the outcome of the binary formation process.
 
The WTTS in US-A and US-B show the same mix in spectral types and
ages. The average Lithium abundance is slightly lower
in US-B compared to US-A. If this is not due to age, then it might
be related to the difference in binary separations. Wider binaries %might
sustain massive circumstellar disks longer than closer binaries. Without a
massive circumstellar disk a T\,Tauri star cannot effectively dissipate angular
momentum and thus spins up while contracting (e.g.\ Bodenheimer 1995).
Faster rotation should enhance the dynamo effect and the
resulting stronger internal magnetic fields could in turn aggravate convection 
and thus slow down the depletion of Lithium on the surface. 

\section{The Distribution of Binary Separations}

%
% Fig. 2a&b 
%
\begin{figure*}[htb]
\hbox{
\centerline{\plotfiddle{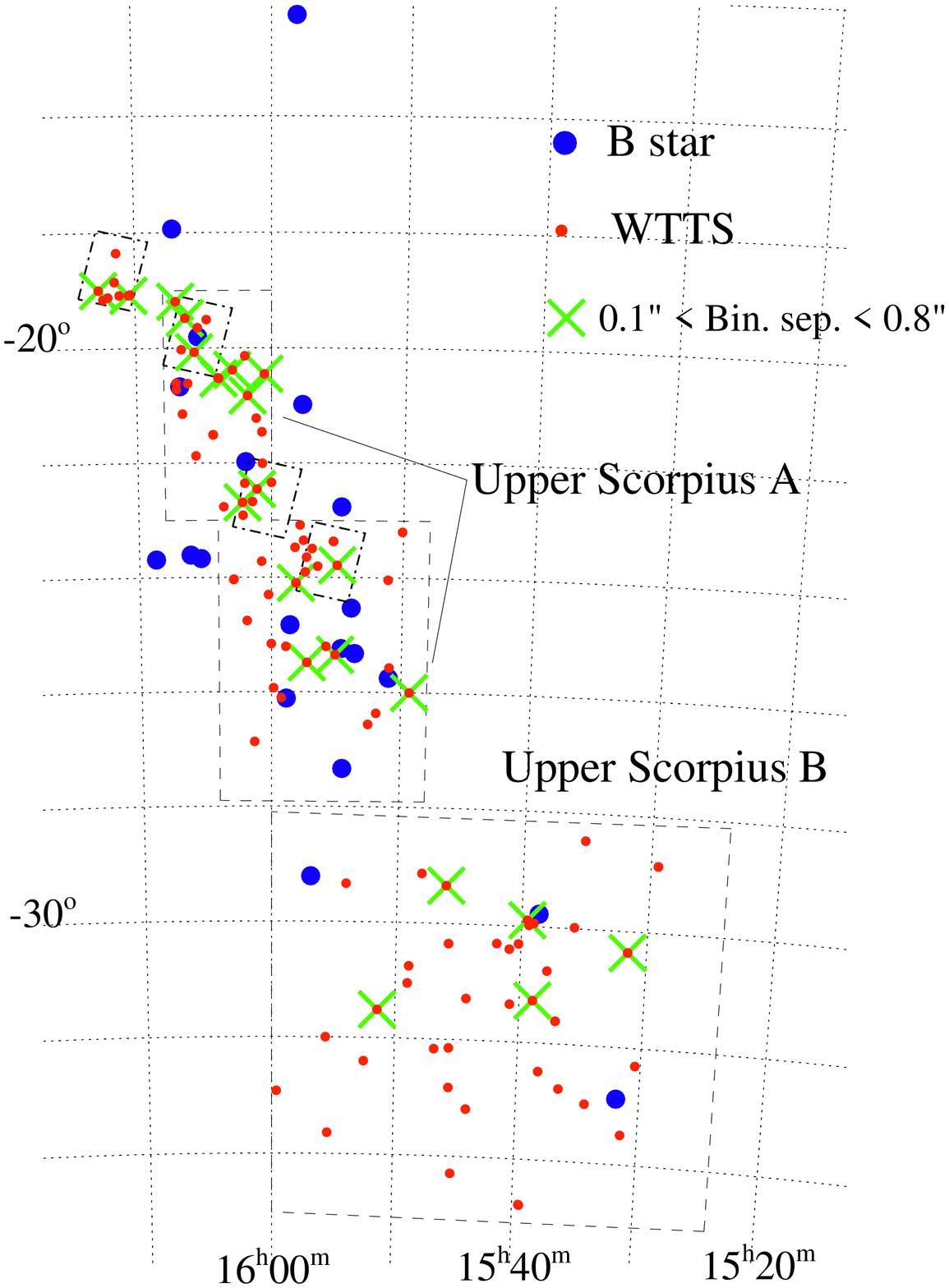}{11cm}{0}{50}{50}{-540}{-50}}
\centerline{\plotfiddle{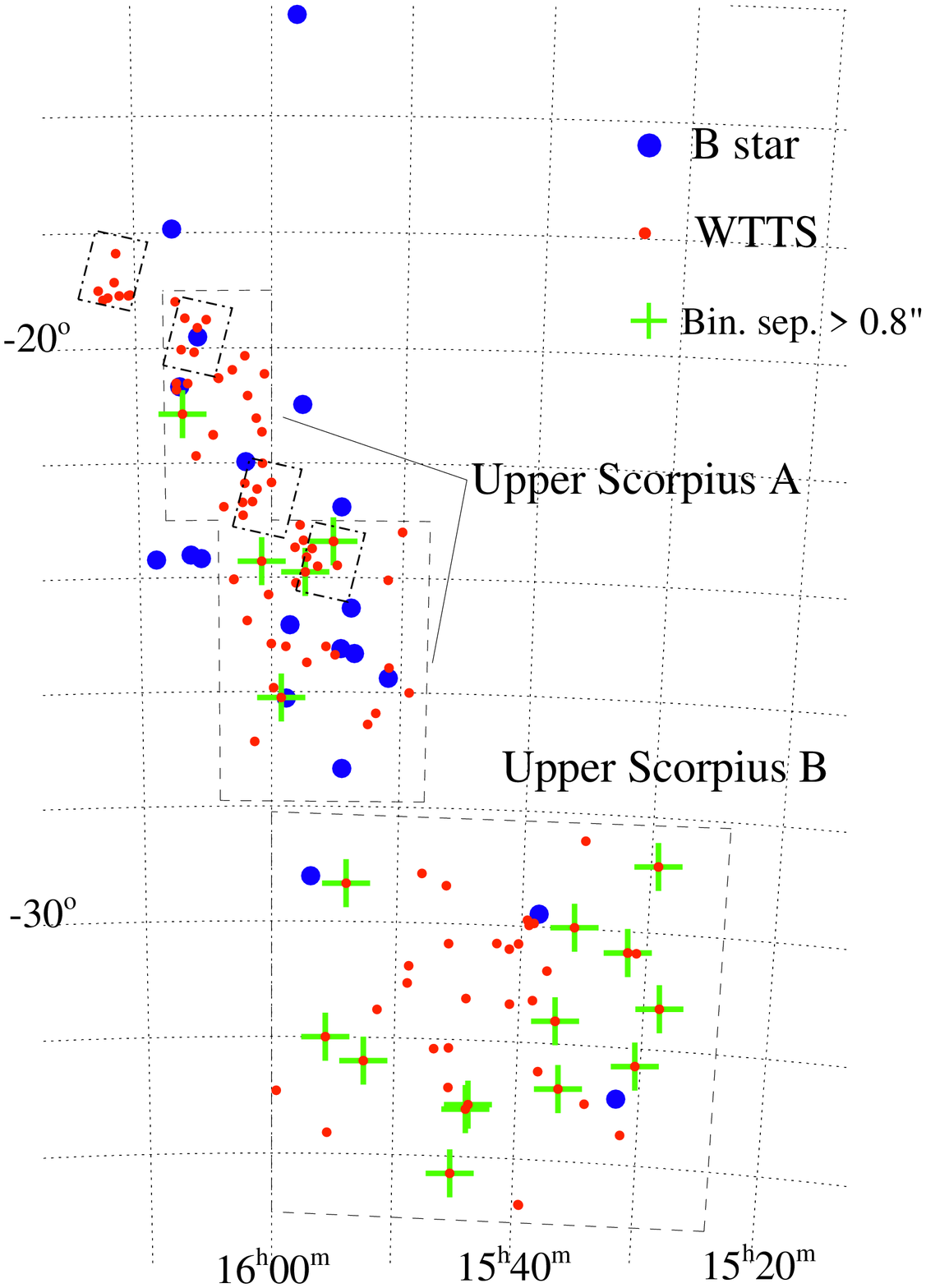}{11cm}{0}{50}{50}{-780}{-50}}}
\figcaption{Close-up of Upper Scorpius A and B.
The regions designated as Upper Scorpius A and B and studied as ROSAT
follow-ups are outlined by dashed lines, the regions observed by the EINSTEIN
satellite are outlined by dash-dotted lines.
The closer binary WTTS cluster in Upper Scorpius A, whereas the wider binaries
are more frequent in Upper Scorpius B.
\label{fig2}}
\end{figure*}

Based on counts of background sources, we expect two of the 39 binaries 
in our sample to be pure chance projections. A 5\% contamination by bogus
binaries should not strongly affect our analysis.
Without correction for selection effects, the `raw' binary frequency
in US-B (39\%$\pm$9\%) is somewhat higher than in US-A (31\%$\pm$7\%),
but the difference is not statistically significant.

The distribution of binary separations, however, is clearly distinct in 
the two regions: about 80\% of the binary systems in US-A have separations less
than 0\farcs8, whereas in US-B 70\% of the binary systems have separations 
greater than 0\farcs8 (Fig.\ 2). The average separation of binary systems in 
US-B is 1\farcs47, more than twice as large as the average separation of binary
systems in US-A (0\farcs63).

%
% Fig.3
%
\begin{figure*}[htb]
\centerline{\plotfiddle{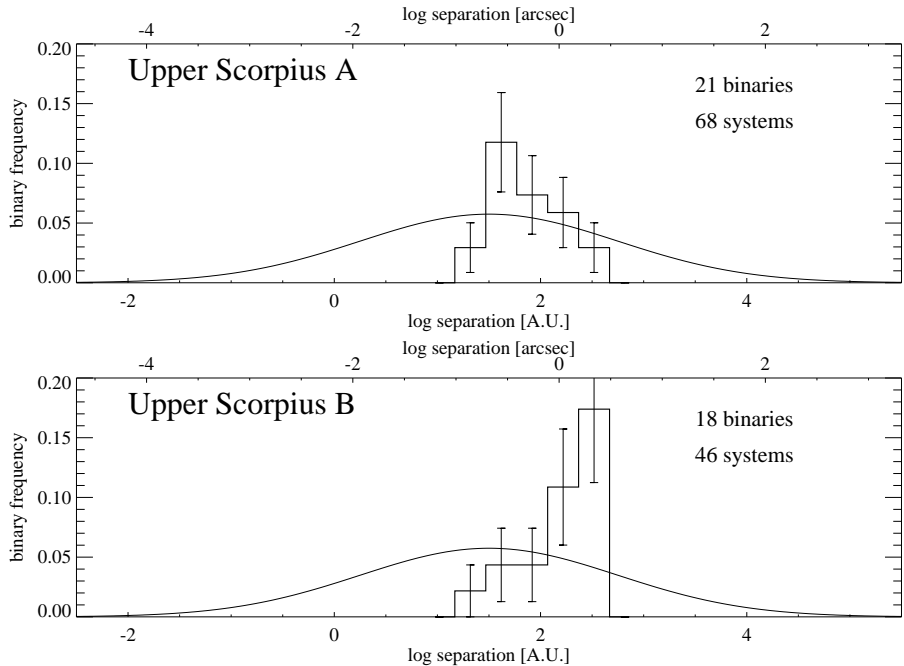}{6.5cm}{90}{50}{50}{-45}{-30}}
\figcaption{Histograms of the distribution of binary star separations for
weak-line T\,Tauri stars in Upper Scorpius A and B. The log-normal type
distribution found by Duquennoy \& Mayor (1991) for main-sequence stars
is shown as well.\label{fig3}}
\end{figure*}

In order to transform angular to physical separations, we need to know
the distance to our stars. The average distance to Scorpius-Centaurus, 
as derived from HIPPARCOS measurements of early type stars, is 145
parsec for US and 140 parsec for UCL (de Zeeuw et al.\ 1998). 
For four of the WTTS in US-B HIPPARCOS derived meaningful distances
(Neuh\"auser \& Brandner 1998; Wichmann et al.\ 1998). One of these
stars (HD 140637) is clearly a foreground star at a distance of
41$^{+2}_{-2}$ parsec. The three other stars (all of them are binaries) have
HIPPARCOS distances between 90 parsec and 160 parsec, albeit with rather large
uncertainties ($\pm$20 to $\pm$80 parsec). We see no evidence that the 
WTTS in US-B are more
than a factor of 2 closer than the WTTS in US-A as would be necessary
in order to explain the difference in the distribution of angular binary
separations by the distance. In the following we assume 
that the WTTS in US-A and US-B are at the same
distance as the early type stars in Scorpius-Centaurus.
Figure 3 shows a histogram of the distribution of binary separations in
US-A and in US-B.
While the distribution peaks around 90 A.U.\ in US-A, the peak of the
distribution is clearly shifted towards larger separations ($\approx$
215 A.U.) in US-B. 
A Kolmogorov-Smirnov test indicates that both binary populations are 
drawn from different distributions at a significance level of 98\%.
Removing the two widest binaries in US-B from the sample (assuming that 
they might be chance projections) lowers the significance level to \%96.

\section{Star forming environments and Binaries}

In Scorpius-Centaurus the molecular cloud in which the star formation originated
has long been dispersed by ionizing radiation and fast stellar winds from the 
massive stars and by supernova explosions. Today, the Lupus dark clouds to the 
south and the $\rho$ Ophiuchi dark clouds to the east of US are the only sites 
of ongoing (low-mass) star formation in the outskirts of a once giant molecular
cloud. No direct trace of the original physical conditions remains. We can, 
however, deduce that the same environment which facilitated the formation of 
massive stars in US-A also facilitated the formation of closer binaries among 
the low-mass stars, whereas in US-B the original physical conditions were such 
that almost no massive stars formed, and preferentially wide binaries among the
low-mass stars.

Other studies might reveal a similar trend. In their search for visual binaries
in nearby dark clouds Reipurth \& Zinnecker (1993) found a weak anti-correlation
between the binary frequency and the number of young stars in the clouds. This 
could be due to a shift in the peak of the distribution of binary separations, 
with the larger clouds producing preferentially closer binaries than the smaller
clouds. Padgett et al.\ (1997) and Petr et al.\ (1998) studied the binary 
frequency in
Orion (distance $\approx$ 460 pc) and found a deficit of ``wide'' 
($\ge$0\farcs1, i.\,e., 50 A.U.) pre-main-sequence binaries among T\,Tauri 
stars closely associated to the high-mass stars in the Trapezium cluster.
 
Binary stars form due to fragmentation during collapse (e.g.\ Boss 1997 and 
references therein). The binary separation depends on the initial angular 
momentum and on the critical density at which the magnetic field 
decouples from the matter (Mouschovias 1977).
The critical density $n_{\rm cr}$ relates to the separation
${\rm a}_{\rm b}$ of a binary systems as
$n_{\rm cr} \propto {\rm a}_{\rm b}^{-3/4}$.
Part of the initial angular momentum of a protobinary might be transported
outwards by means of a circumbinary disk or might be dissipated due to
the gravitational interaction of the circumstellar disks of the individual
components. The temperature structure of a molecular cloud could influence
the number of binary stars (Durisen \& Sterzik 1994) as well as their typical 
separation (Durisen et al., in prep.).
The limited parameter space due to a higher ambient cloud temperature
would preferably lead to the formation of closer binaries. 

%It seems that the distribution of binary separations is closely related
%to the physical conditions in the parental molecular cloud at the time
%star formation sets in. The distribution of binary separations
%provides a fossil record of these conditions. 
Molecular clouds in the process of forming (binary) stars have to be studied
in order to learn which physical quantities determine the outcome of the binary 
formation process.

\section{Conclusion}

We have shown that the distributions of binary separations among weak-line 
T\,Tauri stars in two adjacent fields (`US-A' and `US-B') in the 
Scorpius-Centaurus OB association are clearly distinct from each other
and considerably more peaked than the (broad) distribution of binary 
separations observed among main-sequence field stars. 
In US-A, the WTTS are closely associated with B type stars, whereas
in US-B only a few early type stars are present. We conclude that the
same physical conditions which facilitate the formation of massive
stars also facilitate the formation of closer binaries among low-mass
stars, whereas physical conditions unfavorable for the formation of
massive stars lead to the formation of wider binaries among low-mass
stars.

The outcome of the binary formation process might be determined by the critical
density at which the magnetic
field support breaks down, the internal turbulence and the angular momentum 
of molecular cloud cores, the initial temperature within a cloud, or - most 
likely - a combination of all of these. 

We further conclude that the distribution of binary separations is not a
universal quantity. Instead, both the peak and the width of the 
distribution might vary from one star forming region to the next. 
The broad distribution of binary separations observed among 
main-sequence field stars can be understood as a superposition of
binary populations originating in various star forming environments
with very distinct peaks in the distribution of binary separations.

The apparent overabundance of binaries among T\,Tauri
stars in the Taurus-Auriga T\,association might be explained by the
fact that the distribution of binary separations there is strongly
peaked towards $\approx$ 30\,A.U. Extrapolating from this very pronounced
peak over the whole range of possible binary separations then leads
to an erroneously high estimate of the overall binary frequency.

\acknowledgements{We would like to thank Michael Kunkel for making his
list of WTTS in US available to us prior to publication. We acknowledge
inspiring discussions with Dick Durisen, Christoph Leinert, Michael 
Sterzik, and Hans Zinnecker, who initiated this project. WB acknowledges
support under NASA/HST grant GO-7412.01-94A.}

%\newpage

\end{document}